\documentclass[preprint,review,12pt]{elsarticle}

\usepackage{amssymb}
\usepackage{amsmath,bm}
\usepackage{colortbl,booktabs}
\usepackage{tabularx}
\usepackage{threeparttable}
\usepackage{multicol,multirow}
\usepackage{subfigure}
\usepackage [font=footnotesize,labelfont=bf]{caption}
\usepackage{rotating}
\usepackage{tikz}

\usepackage{color}
\definecolor{R}{rgb}{1, 0, 0}
\definecolor{G}{rgb}{0, 0.5, 0}
\definecolor{B}{rgb}{0, 0, 1}

\definecolor{yjzhuCol}{rgb}{0.9, 0, 0.2}
\definecolor{xyhuCol}{rgb}{0.9, 0, 0.9}

\journal{Journal of Computational Physics}

\begin{document}

\begin{frontmatter}

\title{A L2-norm regularized incremental-stencil WENO scheme for compressible flows}

\author{Yujie Zhu}
\ead{yujie.zhu@tum.de}
\author{Xiangyu Hu \corref{mycorrespondingauthor}}
\cortext[mycorrespondingauthor]{Corresponding author. Tel.: +49 89 289 16152.}
\ead{xiangyu.hu@tum.de}
\address{Department of Mechanical Engineering, Technical University of Munich\\
85748 Graching, Germany}

\begin{abstract}
For the simulation of compressible flow with a broadband of length scales and discontinuities,
the WENO schemes using incremental stencil sizes other than uniform ones are promising for more robustness and less numerical dissipation.
However, in smooth region, large weights may be assigned to smaller stencils
due to the lack of high-order derivatives in the smoothness indicator compared with that of larger stencils, 
and may degrade the order accuracy and lead too much numerical dissipation to resolve fine flow structures.
In order to cope with this drawback, based on the stencil selection of WENO-IS [Wang et al., IJMF 104 (2018): 20-31],
we propose a L2-norm regularized incremental-stencil WENO scheme.
In this method, a new L2-norm regularization is introduced into the WENO weighting strategy to modulate the weights of incremental-width stencils by taking account the L2-norm error term. 
A high-order non-dimensional discontinuity detector is then utilized as the regularization parameter for adaptive control.
In addition, a hybrid method is adopted to further improve the performance and the computational efficiency.
A number of benchmark cases suggest that the present scheme achieves very good robustness and fine-structure resolving capabilities.
\end{abstract}


\end{frontmatter}


\section{Introduction}
\label{sec1}
The simulation of compressible flow is challenging when there present a broadband of length scales and discontinuities.
The numerical schemes must be of low dissipation in smooth region without damping the fine flow structures, 
and be able to introduce sufficient dissipation in vicinity of discontinuities for suppressing non-physical oscillations. 
In order to address this issue, many high-order shock capturing schemes, 
like total variation diminishing (TVD)\cite{harten1983high}, 
essentially non-oscillatory (ENO)\cite{harten1987uniformly} 
and weighted essentially non-oscillatory (WENO)~\cite{liu1994weighted} schemes, have been developed.
Among them, the WENO-JS scheme proposed by Jiang and Shu \cite{jiang1996efficient} exhibits high-order accuracy 
and robust shock-capturing properties 
and have been extensively employed in direct numerical simulations (DNS) and large eddy simulations (LES). 
In spite of these advantages, the classic WENO-JS scheme still faces several shortcomings.
It suffers from order-degeneration, 
e.g. not able to recover the formal order of accuracy near critical points where the first or higher derivatives vanish~\cite{henrick2005mapped},
and is too dissipative for sustaining fine structures in turbulence or aeroacoustic flows~\cite{martin2006bandwidth}. 
Furthermore, the computation may fail when very strong discontinuities exist 
or multiple discontinuities are too close to each other~\cite{gerolymos2009very}. 

In order to overcome the drawbacks of order-degeneration and excessive dissipation 
of the classical WENO-JS scheme, 
one promising approach is to modify the weighting strategy, e.g. the adaptation mechanism, by increasing the weight of the optimal upwind linear scheme in relatively smooth regions.
Following this idea, various schemes have been developed, 
such as WENO-M scheme~\cite{henrick2005mapped} with a corrective mapping applied to the classical WENO weights,  
WENO-Z scheme~\cite{borges2008improved,castro2011high} with a new high-order smoothness indicator
and WENO-RL or WENO-RLTV~\cite{taylor2007optimization} scheme switching on the optimal weights 
when the magnitude of smoothness indicators are comparable. 
Another approach is to introduce the contribution of the downwind stencil and use optimal weights 
for central scheme instead of upwind scheme as pointed in Refs.~\cite{martin2006bandwidth,sun2011class,hu2010adaptive}. 
Take WENO-CU6 scheme~\cite{hu2010adaptive} as an example, Hu et al. employed the 6th-order central scheme as the optimal linear scheme
and introduced a new smooth indicator
for the adaptation between central and upwind stencils.
Later, this scheme was extended to WENO-CU6-M with a physically-motivated scale separation approach for implicit large eddy simulation (ILES)~\cite{hu2011scale}.
While WENO-CU6-M works well for incompressible and compressible turbulence simulation with low dissipation 
and maintains the shock-capturing capability,
it produces spurious waves because of the non-dissipative optimal scheme.
In addition, as demonstrated in Ref.~\cite{fu2016family}, one version of WENO-CU6-M scheme fails in the benchmark test cases "Interacting blast waves" 
and "Double Mach reflection" 
and suggests inferior robustness than the classic WENO-JS scheme. 
Fu et al.~\cite{fu2016family} proposed a family of TENO scheme which dynamically  
assembles a set of low-order candidate stencils with incrementally increasing stencil width. 
The combination of low-order approximation polynomials with incremental stencils 
makes TENO scheme considerably more robust.  
However, positivity-preserving methods, such as those in Ref.~\cite{hu2013positivity,zhang2010positivity}, 
are still required for TENO schemes when high density ratio or strong shock waves present.

Recently, Wang et al.~\cite{wang2018incremental} developed a 5th-order incremental-stencil 
WENO (denote as WENO-IS) scheme for multi-phase flows.
Since both 2- and 3-point stencils are used, 
in the region with strong discontinuities, 
the reconstruction would prefer to choose the 2-point stencils,
which makes the scheme, compared to the classic WENO-JS schemes,
very robust for material interface with high density ratio and strong shock waves.
In their weighting strategy, 
a slight modification is introduced to decrease the weights for 2-point stencils in relatively smooth region
 and to avoid excessive dissipation and order degeneration near critical points.
However,
this scheme is still too dissipative to capture fine flow structures, 
as also shown in Section 4, due to the too much weights for the 2-point stencils.

In this paper, 
we propose a L2-norm regularized incremental-stencil WENO scheme following 
the idea of neural-network optimization~\cite{nielsen2015neural}. 
Similar to WENO-IS scheme, both incremental 2- and 3-point stencils are used for the 5th-order reconstruction.
The difference is that a L2-norm regularization is introduced to address the problem of order degeneration and excessive dissipation. 
First, the L2-norm errors between the approximations from 2-point stencils and a classical 3-point stencil are utilized to modulate the weights of the incremental stencils.  
Then, a high-order non-dimensional discontinuity detector is adopted as the regularization parameter to control the influence of the error term. 
With a further hybridization with a linear scheme, less numerical dissipation in very smooth region and considerable higher computational efficiency are achieved.
The reminder of the paper is organized as follows. 
Section 2 gives a brief review of the WENO-IS scheme following Wang et al.~\cite{wang2018incremental} 
and the application of WENO-IS to Euler equations is given.
In Section 3, the L2-norm regularization 
and control parameter are proposed. 
Benchmark test cases with strong discontinuities and fine flow structures are presented in Section 4, 
and brief concluding remarks are given in the last Section 5.

\section{WENO-IS scheme}
\label{sec1}
In this section, we briefly review the 5th-order WENO-IS scheme for hyperbolic conservation laws 
following Ref.~\cite{wang2018incremental}. 
First, we illustrate the details of WENO-IS scheme based on the discretization of one-dimensional scalar hyperbolic conservation law.
Then, its application in Euler equations is given.
\subsection{Semi-discretization of a hyperbolic conservation law}
\label{subsec1}
For simplicity, we consider the following one-dimensional scalar hyperbolic conservation law
 \begin{equation}
 \frac{\partial {u}}{{\partial t}} +
 \frac{\partial {f\left({u} \right) }}{{\partial x}}=0,
 \label{goveq3}
 \end{equation}
 where $u$ denotes the conservative variable and $f\left(u \right) $ is the flux function.
 Here, the characteristic velocity is assumed to be positive with $\frac{\partial {f\left({u} \right) }}{{\partial u}}>0$.
 
 On a uniform grid, we denote $x_j=j\Delta x$, where $\Delta x$ is the grid spacing. 
 The quantities at the point $x_j$ are identified by the subscript $j$.
 The semi-discretized form of Eq.~\eqref{goveq3} using the method of lines can be 
 approximated by a conservative finite difference scheme as  
 \begin{equation}
\frac{d {{u}_j}}{{d t}} =
-\frac{1}{\Delta x} \left(h_{j+1/2}-h_{j-1/2} \right) ,
\label{eq2}
\end{equation}
 where $h_{j\pm1/2}=h\left( x_{j\pm1/2}\right) $ and is implicitly defined by 
  \begin{equation}
 f\left[u\left( x\right)  \right]  =
 \frac{1}{\Delta x} \int_{x-\Delta x/2}^{x+\Delta x/2} h(\xi)d\xi.
 \label{eq3}
 \end{equation}
 Numerically, $h_{j\pm1/2}$ are approximated by high-order polynomial interpolations $\widehat{f}_{j\pm1/2}$ or numerical fluxes and Eq.~\eqref{eq2} can be expressed as 
  \begin{equation}
 \frac{d {u_j}}{{d t}} \approx
 -\frac{1}{\Delta x} \left(\widehat{f}_{j+1/2}-\widehat{f}_{j-1/2} \right).
 \label{eq4}
 \end{equation}

\subsection{Candidate stencils with incremental width}
\label{subsec1}
\begin{figure}[tbh]
 	\begin{center}
 		\begin{tikzpicture}[
 		dot/.style 2 args={circle,draw=#1,fill=#2,inner sep=2.5pt},
 		square/.style 2 args={draw=#1,fill=#2,inner sep=3pt},
 		mystar/.style 2 args={star,draw=#1,fill=#2,inner sep=1.5pt},
 		mydiamond/.style 2 args={diamond,draw=#1,fill=#2,inner sep=1.5pt},
 		scale=0.8
 		]
 		
 		\draw[black,thick]  (0,-1) grid (15,-1);
 		\draw[black,thick,yshift=-0.3cm]  (6.25,-3) grid (8.75,-3);
 		\draw[black,thick,yshift=-0.3cm]  (3.75,-4) grid (6.25,-4);
 		\draw[black,thick,yshift=-0.3cm]  (6.25,-5) grid (11.25,-5);
 		\draw[black,thick,yshift=-0.3cm]  (1.25,-6) grid (6.25,-6);
 		\draw[gray,thick,yshift=-0.3cm]  (3.75,-7) grid (8.75,-7);
 		\draw[black,thick,yshift=-0.3cm]  (1.25,-8) grid (11.25,-8);
 		\foreach \Fila in {1.25,3.75,6.25,8.75,11.25,13.75}{\node[dot={black}{black}] at (\Fila,-1) {};}   
 		\foreach \Fila in {7.5}{\node[square={black}{white}] at (\Fila,-1) {};}  
 		\foreach \Fila in {6.25,8.75}{\node[dot={black}{black}] at (\Fila,-3.3) {};}   
 		\foreach \Fila in {3.75,6.25}{\node[dot={black}{black}] at (\Fila,-4.3) {};}  
 		\foreach \Fila in {6.25,8.75,11.25}{\node[dot={black}{black}] at (\Fila,-5.3) {};}   
 		\foreach \Fila in {1.25,3.75,6.25}{\node[dot={black}{black}] at (\Fila,-6.3) {};}  
 		\foreach \Fila in {3.75,6.25,8.75}{\node[dot={gray}{gray}] at (\Fila,-7.3) {};} 
 		\foreach \Fila in {1.25,3.75,6.25,8.75,11.25}{\node[dot={black}{black}] at (\Fila,-8.3) {};} 
 		\node[below] at (1.25,-1.3) {$x_{j-2}$};
 		\node[below] at (3.75,-1.3) {$x_{j-1}$};
 		\node[below] at (6.25,-1.3) {$x_{j}$};
 		\node[below] at (8.75,-1.3) {$x_{j+1}$};
 		\node[below] at (11.25,-1.3) {$x_{j+2}$};
 		\node[below] at (13.75,-1.3) {$x_{j+3}$};
 		\node[below] at (7.5,-1.3) {$x_{j+1/2}$};
 		\node[left]  at (6.25,-3.3) {$S_{0}$};
 		\node[left]  at (3.75,-4.3) {$S_{1}$};
 		\node[left]  at (6.25,-5.3) {$S_{2}$};
 		\node[left]  at (1.25,-6.3) {$S_{3}$};
 		\node[left]  at (3.75,-7.3) {$S_{01}$};
 		\node[left]  at (1.25,-8.3) {$S_{5}$};
 		\end{tikzpicture}
 	\end{center}		
 	\caption{Full stencil and candidate stencils with incremental width for WENO-IS reconstruction. 
 	Here, $S_{01}$ is an original 3-point stencil used for WENO-JS reconstruction.  }
 	\label{stencil}	
 \end{figure}
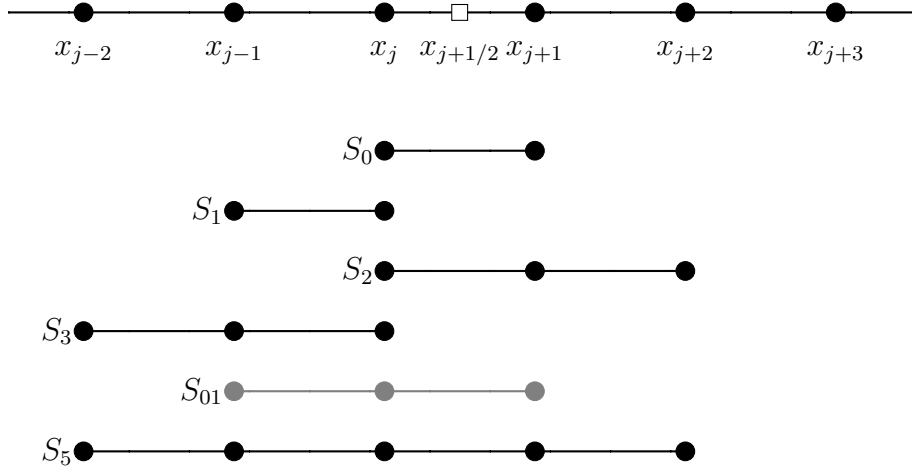

 For 5th-order WENO-IS scheme, the 5-point stencil $S_5$ as shown in Fig.~\ref{stencil} is served as 
 the full stencil for the evaluation of numerical fluxes $\widehat{f}_{j+1/2}$.
 Different from the classic WENO-JS scheme, which 
 subdivides the full stencil into three stencils with the same width, 
 the WENO-IS scheme uses four stencils with incremental width
 including two 2-point and two 3-point stencils, e.g. $S_0$, $S_1$, $S_2$ and $S_3$ 
 as shown in Fig.~\ref{stencil}, for reconstruction. 
 
 The numerical flux $\widehat{f}_{j +1/2} $ is computed by a 
 convex combination of the four candidate stencils fluxes as
\begin{equation}
\widehat{f}_{j +1/2}=
\sum_{k=0}^{3} \omega_k \widehat{f}_{k,j+1/2} ,
\label{eq5}
\end{equation}
where $\widehat{f}_{k,j+1/2}$ and $\omega_k$ are the reconstructed fluxes from candidate stencils 
and their non-linear weights, respectively.
These reconstructed fluxes are
\begin{equation}
\begin{split}
&\widehat{f}_{0,j+1/2}=\frac{1}{2}f_j+\frac{1}{2}f_{j+1}, \\
&\widehat{f}_{1,j+1/2}=-\frac{1}{2}f_{j-1}+\frac{3}{2}f_{j} , \\
&\widehat{f}_{2,j+1/2}=\frac{1}{3}f_{j}+\frac{5}{6}f_{j+1}-\frac{1}{6}f_{j+2} , \\
&\widehat{f}_{3,j+1/2}=\frac{1}{3}f_{j-2}-\frac{7}{6}f_{j-1}+\frac{11}{6}f_{j}.  \\
\end{split} 
\label{eq6}
\end{equation}
From Taylor series expansion, we can obtain
\begin{equation}
\widehat{f}_{k,j +1/2}=
h_{j+1/2}+A_k \Delta x^{r_k}+O\left( \Delta x^{r_k+1}\right)  ,
\label{eq7}
\end{equation}
where $A_k$, $k=0, 1, 2, 3$ are independent of $\Delta x$ and $r_k$ denote the points of the candidate stencil.

The non-linear weights $\omega_k$ are given by
\begin{equation}
\omega_k=
\frac{\alpha_k}{\sum\nolimits_{s=0}^{3}\alpha_s} , \quad \alpha_k=d_k\left(1+\frac{\tau_5}{\beta_k+\varepsilon} \right), 
\label{eq8}
\end{equation}
where coefficients $d_0=\frac{4}{10}, d_1=\frac{2}{10}, d_2=\frac{3}{10}, d_3=\frac{1}{10}$ are the optimal weights 
and $\varepsilon$ is a small positive parameter to prevent division by zero and we choose $\varepsilon=10^{-20}$ in this paper.
 $\beta_k$ are the smoothness indicators of each candidate stencil, 
while $\tau_5$ is the global reference smoothness indicator on the full stencil. Following Jiang and Shu ~\cite{jiang1996efficient},
the smoothness indicators at $ S_k$ are evaluated by
 \begin{equation}
 \beta_k=\sum\limits_{l=1}^{r_k-1}\Delta x^{2l-1} \int_{x_{j-1/2}}^{x_{j+1/2}}\left(\frac{d^l}{dx^l} 
 \widehat{f}_k\left( x\right) \right)^2 dx. 
 \label{eq9}
 \end{equation}
The global reference smoothness indicator is obtained by
\begin{equation}
\tau_5=
\sum\limits_{l=3}^{4}\Delta x^{2l-1} \int_{x_{j-1/2}}^{x_{j+1/2}}\left(\frac{d^l}{dx^l} \widehat{f}_k\left( x\right) \right)^2 dx,  
\label{eq10}
\end{equation}
which only contains the high-order component of the smoothness indicator for the full stencil~\cite{fu2016family,wang2018incremental}.
The formulations of $\beta_k$ and $\tau_5$ in terms of cell-centered fluxes $f_j$ can be expressed as
\begin{equation}
\begin{split}
&\beta_0=\left(f_{j+1}-f_{j} \right) ^2, \\
&\beta_1=\left(f_{j}-f_{j-1} \right) ^2, \\
&\beta_2=\frac{13}{12}\left(f_{j}-2f_{j+1}+f_{j+2} \right) ^2+ \frac{1}{4}\left(3f_{j}-4f_{j+1}+f_{j+2} \right) ^2, \\
&\beta_3=\frac{13}{12}\left(f_{j-2}-2f_{j-1}+f_{j} \right) ^2+ \frac{1}{4}\left(f_{j-2}-4f_{j-1}+3f_{j} \right) ^2, \\
&\tau_5=\frac{13}{12}\left(f_{j+2}-4f_{j+1}+6f_{j}-4f_{j-1}+f_{j-2} \right) ^2+ 
\frac{1}{4}\left(f_{j+2}-2f_{j+1}+2f_{j-1}-f_{j-2} \right) ^2, \\
\end{split} 
\label{eq11}
\end{equation}
and their Taylor series expansions at $x_j$ are
\begin{equation}
\begin{split}
&\beta_0={f^\prime}_j ^2 \Delta x^2+{f^\prime}_j{f^\prime{^\prime}}_j \Delta x^3 +
\left(\frac{1}{4}{f^\prime{^\prime}}_j^2+\frac{1}{3}{f^\prime}_j{f^\prime{^\prime}{^\prime}}_j \right)\Delta x^4+
\left(\frac{1}{12}{f^\prime}_j{f^\prime{^\prime}{^\prime}{^\prime}}_j+
\frac{1}{6}{f^\prime{^\prime}}_j{f^\prime{^\prime}{^\prime}}_j \right)\Delta x^5+ 
O\left( \Delta x^6\right) , \\
&\beta_1={f^\prime}_j ^2 \Delta x^2-{f^\prime}_j{f^\prime{^\prime}}_j \Delta x^3 +
\left(\frac{1}{4}{f^\prime{^\prime}}_j^2+\frac{1}{3}{f^\prime}_j{f^\prime{^\prime}{^\prime}}_j \right)\Delta x^4-
\left(\frac{1}{12}{f^\prime}_j{f^\prime{^\prime}{^\prime}{^\prime}}_j+
\frac{1}{6}{f^\prime{^\prime}}_j{f^\prime{^\prime}{^\prime}}_j \right)\Delta x^5+ 
O\left( \Delta x^6\right) , \\
&\beta_2={f^\prime}_j ^2 \Delta x^2+
\left(\frac{13}{12}{f^\prime{^\prime}}_j^2-\frac{2}{3}{f^\prime}_j{f^\prime{^\prime}{^\prime}}_j \right)\Delta x^4+
\left(-\frac{1}{2}{f^\prime}_j{f^\prime{^\prime}{^\prime}{^\prime}}_j+
\frac{13}{6}{f^\prime{^\prime}}_j{f^\prime{^\prime}{^\prime}}_j \right)\Delta x^5+ 
O\left( \Delta x^6\right) , \\
&\beta_3={f^\prime}_j ^2 \Delta x^2+
\left(\frac{13}{12}{f^\prime{^\prime}}_j^2-\frac{2}{3}{f^\prime}_j{f^\prime{^\prime}{^\prime}}_j \right)\Delta x^4-
\left(-\frac{1}{2}{f^\prime}_j{f^\prime{^\prime}{^\prime}{^\prime}}_j+
\frac{13}{6}{f^\prime{^\prime}}_j{f^\prime{^\prime}{^\prime}}_j \right)\Delta x^5+ 
O\left( \Delta x^6\right) , \\
&\tau_5={f^\prime{^\prime}{^\prime}}_j\Delta x^6+\frac{13}{12}{f^\prime{^\prime}{^\prime}{^\prime}}_j\Delta x^8 
+ O\left( \Delta x^{10}\right). \\
\end{split} 
\label{eq12}
\end{equation}
Substituting Eq.~\eqref{eq12} to Eq.~\eqref{eq8}, we can observe that 
\begin{equation}
\frac{\tau_5}{\beta_k+\varepsilon}=O\left(\Delta x^4 \right)  , \quad k=0,1,2,3.
\label{eq13}
\end{equation}
Therefore, the condition of $\omega_k-d_k=O\left(\Delta x^4 \right)$ is satisfied at the region without critical points,
 implying that
the WENO-IS scheme has the expected 5th-order convergence from the analysis in Ref.~\cite{wang2018incremental}.

\subsection{Weighting strategy}
\label{subsec1}
From Eq.~\eqref{eq9} and Eq.~\eqref{eq11}, we can observe that the smoothness indicators of 2-point stencils 
are evaluated from the 1st-order derivatives of linear functions, 
while these of 3-point stencils are from the 1st and 2nd-order derivatives of parabolic functions.
Due to the lack of 2nd-order derivative term, the smoothness indicator of 2-point stencils can be much smaller 
than that of stencils with 3-point especially near critical point and larger weights will be posed on 2-point stencils, 
which may cause order degeneration eventually. In order to address this issue, 
Wang et al.~\cite{wang2018incremental} made a slight modification of the weighting strategy in Eq.~\eqref{eq8} as
\begin{equation}
\alpha_0=d_0\left(1+\frac{\tau_5}{\beta_0+\varepsilon }\cdot \frac{\tau_5}{\beta_{01}+\varepsilon } \right), \quad 
\alpha_1=d_1\left(1+\frac{\tau_5}{\beta_1+\varepsilon }\cdot \frac{\tau_5}{\beta_{01}+\varepsilon } \right),
\label{eq14}
\end{equation}
where 
\begin{equation}
\beta_{01}=\frac{13}{12}\left(f_{j-1}-2f_{j}+f_{j+1} \right) ^2+ \frac{1}{4}\left(f_{j-1}-f_{j+1}\right) ^2
\label{eq15}
\end{equation}
is the smoothness indicator of $S_{01}$, an original 3-point stencil of WENO-JS as shown in Fig.~\ref{stencil}, 
and is obtained from 1st and 2nd-order derivatives. 
This modification decreases the weights of 2-point stencils considerably and avoids order degeneration at critical point.

In relatively smooth region, the effectiveness of this modification 
has been confirmed by a number of test cases in Ref.~\cite{wang2018incremental}.
The WENO-IS scheme achieves the formal order of accuracy and produces less errors than 
classical WENO-JS scheme.
However, in the region with shocks or high wavenumber waves, 
this modification may not decrease the weights of 2-point stencils sufficiently,
which can still lead  to excessive numerical dissipation. 
Suppose there is a shock locating in stencil $S_3$ and the solutions in 
stencils $S_0$, $S_1$ and $S_2$ are smooth (see Fig.~\ref{stencil}).
Under this condition, $\tau_5$ is comparable to $\beta_3$ and $\tau_5=O\left( 1\right) $, 
while $\beta_k=O\left(\Delta x^2 \right)  $(k=0,1,2,01) are all small numbers. 
It is straightforward to show that
\begin{equation}
\frac{\tau_5}{\beta_0+\varepsilon }\cdot \frac{\tau_5}{\beta_{01}+\varepsilon } \gg \frac{\tau_5}{\beta_{2}+\varepsilon }, \quad
\frac{\tau_5}{\beta_1+\varepsilon }\cdot \frac{\tau_5}{\beta_{01}+\varepsilon } \gg \frac{\tau_5}{\beta_{2}+\varepsilon },
\label{eq144}
\end{equation}
which means larger weights will be posed on 2-point stencils.
Same conclusion can be drawn when the shock locates in other stencils.
Thus, this modification enlarges the numerical dissipation near the discontinuities.
Although larger numerical dissipation improves the robustness of the scheme,
this modification imposes too much excessive numerical dissipation to resolve the fine flow structures with broadband length scales well (also see the cases in section 4).  
 
Note that TENO scheme ~\cite{fu2016family} avoid the problem of large weight posed on low-order stencils
 by a ENO-like strategy.
However, the numerical stability of TENO scheme is inferior to WENO-JS hence also WENO-IS scheme. 

\subsection{Application in Euler equations}
\label{subsec1}
For simplicity, the detailed procedure on the application of WENO-IS for Euler equations is illustrated based on 
the one-dimensional Euler equations of gas dynamics
\begin{equation}
\frac{\partial \bm{U}}{{\partial t}} +
\frac{\partial {\bm{F}\left(\bm{U} \right) }}{{\partial x}}=0,
\label{goveq4}
\end{equation}
where 
\begin{equation}
\bm{U}=\left[\begin{array}{c}
\rho \\ \rho u \\ {E}
\end{array} \right] , 
\bm{F}\left(\bm{ U}\right) =\left[\begin{array}{c}
\rho u \\ \rho u^2+p \\ {\left( E+p\right)u }
\end{array} \right]. \\
\end{equation}
Here, $\rho$ is density, $p$ is pressure and $u$ denotes the velocity in $x-$direction.
$E=p/\left( \gamma-1\right) +1/2\rho u^2$ is total energy per unit volume with ideal-gas equation of states used. 

Within a uniform grid, at the point $x_j$, the semi-discretized form of Eq. \eqref{goveq4} gives
\begin{equation}
\frac{d {\bm{U}_j}}{{d t}} =-
\frac{1}{\Delta x}\left(\hat{\bm{F}}_{j+1/2} -\hat{\bm{F}}_{j-1/2}\right) ,
\label{goveq5}
\end{equation}
where $\hat{\bm{F}}_{j\pm1/2}$ are the numerical fluxes at $x_{j\pm 1/2}$. 
With a typical characteristic-wise method, the numerical fluxes
are reconstructed within the local characteristic flied. 
The procedure to obtain $\hat{\bm{F}}_{j+1/2}$ includes the following steps:

(1) At $x_{j+ 1/2}$, evaluate the Jacobian matrix $\bm{A}=\partial \bm{F}/\bm{U}$ at a Roe-average state.

(2) Compute the left and right eigenvector matrix $\bm{L}_{j+1/2}^s$, $\bm{R}_{j+1/2}^s \left(s=1,2,3 \right) $ 
as well as the eigenvalues $\lambda_{j+1/2}^s \left(s=1,2,3 \right)$ 
of the Roe-average Jacobian matrix $\bm{A}$ .

(3) Within each candidate stencils, transform the physical fluxes and conservative variables into characteristic filed as 
\begin{equation}
v_m^s =\bm{L}_{j+1/2}^s \cdot \bm{U}_m, \quad
g_m^s =\bm{L}_{j+1/2}^s \cdot \bm{F}_m, 
\label{eqchara}
\end{equation}
where $j-2<m<j+3$ and $s=1,2,3$.

(4) Carry out flux splitting in characteristic space. We can obtain
\begin{equation}
f_m^{s,\pm} =\frac{1}{2} \left(g_m^s \pm \alpha^s v_m^s \right) , 
\label{flxuspliting}
\end{equation}
where $\alpha^s=\left|\lambda_{j+1/2}^s \right| $ for Roe flux (RF) splitting. 
Otherwise, $\alpha^s=max \left|\lambda_{l}^s \right| $ for Lax-Friedrichs flux (LF) splitting with
 $l$ representing the entire computational domain or for local Lax-Friedrichs flux (LLF) splitting when $l$ represents the local stencil.
 
 (5) Then, reconstruct the numerical flux with WENO-IS scheme, which gives
 \begin{equation}
 f_{j+1/2}^{s,+} =\sum\limits_{k=0}^{3} \omega_k^{+}  f_{k,j+1/2}^{s,+} , \quad
 f_{j+1/2}^{s,-} =\sum\limits_{k=0}^{3} \omega_k^{-}  f_{k,j+1/2}^{s,-} ,
 \label{reconstruction}
 \end{equation}
where $f_{k,j+1/2}^{s,+}$ and $\omega_k^{+}$ are the positive fluxes by candidate stencils 
and their non-linear weights as defined in Eq.~\eqref{eq6} and Eq.~\eqref{eq8}, respectively.
The negative term  $f_{k,j+1/2}^{s,-}$ and $\omega_k^{-}$ can be computed in the similar way according to the symmetry at the point $x_{j+1/2}$.
The numerical flux in each characteristic space can be finally evaluated by
 \begin{equation}
\hat{f}_{j+1/2}^{s} =f_{j+1/2}^{s,+}+f_{j+1/2}^{s,-} . 
\label{reconstruction1}
\end{equation}

(6) At last, transform the numerical flux back into physical space as
 \begin{equation}
\hat{\bm{F}}_{j+1/2}  = \sum\limits_{s=0}^{2} \bm{R}_{j+1/2}^{s} \hat{f}_{j+1/2}^{s}. 
\label{reconstruction2}
\end{equation}

\section{A L2-norm regularized WENO-IS scheme}
\label{sec1}
In Ref.~\cite{nielsen2015neural} (their section 3.2.1), 
in order to find a way to reduce overfitting of the neural-network, 
an optimization technique, known as L2-norm regularization, is introduced.
The essential idea of L2-norm regularization is to add an extra term, called the regularization term, 
to the original cost function which only contains the L2-norm error term, with a hyper-parameter to control the relative amount of regularization. 
This technique can be viewed as a way of compromising between the original L2-norm error term and the L2-norm regularization term,
which makes the neural-network better at generalizing beyond the training data.

In the classical WENO schemes, as an analog of optimization technique, we can find that only the regularization terms based on the smoothness indicators are utilized. 
The L2-norm error term is neglected due to the same order of accuracy for each candidate stencil.
However, for WENO-IS scheme, the reconstruction accuracy of 2-point stencil is different from that of the 3-point stencil, therefore,
neglecting the L2-norm error terms is not suitable anymore.
To cope with the discrepancy of numerical accuracy between 2-point and 3-point stencils,
an optimization strategy, L2-norm regularization, is introduced to the WENO weighting strategy.

\subsection{L2-norm regularization}
\label{subsec1}

For the conservation law in Eq.~\eqref{goveq3}, 
a form of L2-norm error term is
 \begin{equation}
E_k=\frac{1}{ \Delta x} \int_{x_{j-1/2}}^{x_{j+1/2}}\left( \widehat{f}_k\left( x\right)-{f}_k\left( x\right) \right)^2 dx, 
\label{eq16}
\end{equation}
which is obtained by the difference between reconstruct flux and reference flux.
For the 5-th order WENO-IS scheme,
the three reconstructed fluxes of classic WENO-JS scheme are chosen as the reference fluxes. 
Therefore, the L2-norm error term of each stencil for WENO-IS scheme can be evaluated by
\begin{equation}
\begin{split}
&E_0=\frac{1}{ \Delta x} \int_{x_{j-1/2}}^{x_{j+1/2}}
\left( \widehat{f}_0\left( x\right)-{f}_{01}\left( x\right) \right)^2 dx, \\
&E_1=\frac{1}{ \Delta x} \int_{x_{j-1/2}}^{x_{j+1/2}}
\left( \widehat{f}_1\left( x\right)-{f}_{01}\left( x\right) \right)^2 dx, \\
&E_2=\frac{1}{ \Delta x} \int_{x_{j-1/2}}^{x_{j+1/2}}\left( \widehat{f}_2\left( x\right)-{f}_2\left( x\right) \right)^2 dx, \\
&E_3=\frac{1}{ \Delta x} \int_{x_{j-1/2}}^{x_{j+1/2}}\left( \widehat{f}_3\left( x\right)-{f}_3\left( x\right) \right)^2 dx, \\
\end{split} 
\label{eq17}
\end{equation}
where $f_{01}\left( x\right)$, $f_{2}\left( x\right)$ and $f_{3}\left( x\right)$ are the reconstructed 
fluxes of the corresponding stencils $S_{01}$, $S_2$ and $S_3$ as shown in Fig.~\ref{stencil}.
Since the reconstructed fluxes and the reference fluxes of stencil $S_2$ and $S_3$ are the same, 
their L2-norm error terms equal to zero accordingly. Therefore, the above error terms yield
\begin{equation}
\begin{split}
&E_0=\frac{1}{45}\left( f_{j+1}^2-4f_j f_{j+1}+2f_{j-1} f_{j+1}+4f_j^2 -4f_{j-1} f_{j}+f_{j-1}^2\right) , \\
&E_1=\frac{1}{45}\left( f_{j+1}^2-4f_j f_{j+1}+2f_{j-1} f_{j+1}+4f_j^2 -4f_{j-1} f_{j}+f_{j-1}^2\right), \\
&E_2=0 , \\
&E_3=0 . \\
\end{split} 
\label{eq18}
\end{equation}

With L2-norm regularization, the weighting strategy of WENO-IS scheme is modified as
\begin{equation}
\omega_k^r=
\frac{\alpha_k^r}{\sum\nolimits_{s=0}^{3}\alpha_s^r} , \quad \alpha_k^r=d_k\left(1+\frac{\lambda\tau_5}{\lambda\beta_k+E_k+\varepsilon} \right), 
\label{eq8-r}
\end{equation}
where $\lambda$ is known as the regularization parameter and we will discuss it in later subsection.
The Taylor series expansions at $x_j$ of the L2-norm error term $E_k$ as shown in Eq.~\eqref{eq18} are
\begin{equation}
\begin{split}
&E_0=E_1=\frac{1}{45}\left( {f^{\prime \prime}}_j \Delta x^4+\frac{1}{6}{f^{\prime \prime}}_j 
{f^{\prime \prime\prime \prime}}_j \Delta x^6+O\left( \Delta x^6\right)\right) , \\
&E_2=E_3=0. \\
\end{split} 
\label{eq277}
\end{equation}
It is straightforward to check from Eq.~\eqref{eq12} that $\omega_k^r-d_k=O\left(\Delta x^4 \right)$ is still satisfied for the present weighting strategy.
Hence, with L2-norm regularization the modified WENO-IS scheme also has the expect 5th-order convergence at non-critical points. 
Note that, this modification doesn't solve the order degeneration problem at critical points.
Although some weighting strategies can be applied to cope with this case as noted by Borges et al.~\cite{borges2008improved}, 
it will increase numerical dissipation inevitably. Alternatively, in the present paper, 
we introduce a hybrid scheme to resolve the
order degeneration problem, which will be discussed in section 3.3.

\subsection{An adaptive weights control }
\label{subsec1}
It is obvious that the effect of the regularization is to 
make up the discrepancy of numerical accuracy between 2-point and 3-points stencils
so as to reduce the weights of 2-point stencils. 
The relative importance of the regularization term and the added L2-norm error term depends on the value of $\lambda$.
Here, we introduce a non-dimensional discontinuity detector following Hu et al.~\cite{hu2015efficient} 
to devise an adaptive regularization parameter.
Due to the design of non-dimensional discontinuity detector is based on Euler equations, in this section, 
the detailed procedure is illustrated using one dimensional Euler equations as presented in section 2.3.

The non-dimensional discontinuity detector is defined by 
\begin{equation}
\sigma^s=\left(\frac{\Delta v_{j+\frac{1}{2}}^s}{\tilde{\rho}} \right) ^{2}, \\
\label{eq27}
\end{equation}
where 
\begin{equation}
\Delta v_{j+\frac{1}{2}}^s=\frac{1}{60} \bm{L}_{j+1/2}^s \cdot 
\left(\bm{U}_{j-2}-5\bm{U}_{j-1}+10\bm{U}_{j}-10\bm{U}_{j+1}+5\bm{U}_{j+2}-\bm{U}_{j+3} \right) , \\
\label{eq28}
\end{equation}
and ${\tilde{\rho}}$ is the Roe-average density corresponding to the Jacobian matrix $\bm{A}$ at $x_{j+\frac{1}{2}}$.

We define a regularization parameter by 
\begin{equation}
\lambda=c\sigma^s, \\
\label{eq29}
\end{equation}
where $c$ is a positive constant and we set $c=1$ in this work.
When the full 6-point stencil contains discontinuities, 
the regularization parameter becomes a large number,
which makes the original smoothness indicator the leading term 
and 2-point stencils will be assigned larger weights.
 Note that, the relatively large weights of 2-point stencils evaluated by the present method 
 are less than these of the original WENO-IS scheme under this condition.
 As will be shown in section 4, with this treatment the scheme is still robust 
 even for strong discontinuities and less numerical dissipation is imposed.
 Otherwise, the reconstruction prefer to choose 3-point stencils in the flow field 
 with fine flow structures or the smooth region where $\lambda$ is small.

\subsection{The hybrid scheme }
\label{subsec1}
Although the added L2-norm error term and the designed adaptive parameter have the advantages as discussed above, 
the modified WENO-IS scheme is not suitable for smooth region due to the preference of 3-point stencil of 
the reconstruction when $\lambda$ is very small. 
Furthermore, compared with its optimal linear scheme, the calculation of the modified WENO-IS scheme is time-consuming.
To overcome these limitations, with the discontinuity detector presented in above section, 
it is easy to introduce a hybrid scheme flowing Hu et al.~\cite{hu2015efficient} to achieve better accuracy in the smooth region 
and improve computational efficiency.
The numerical flux of this hybrid scheme is switching from that of the nonlinear WENO-IS scheme and 
its optimal linear upwind scheme, which is given by
\begin{equation}
\hat{\bm{F}}_{j+1/2}=\sigma_{j+1/2} \hat{\bm{F}}_{j+1/2}^{UPS}+\left(1- \sigma_{j+1/2} \right) \hat{\bm{F}}_{j+1/2}^{WENO-IS} , \\
\label{eqhy}
\end{equation}
where $\sigma_{j+1/2}$ equals to one in smooth region otherwise zero in vicinity of discontinuities. 
Here, the numerical flux $\hat{\bm{F}}_{j+1/2}^{UPS}$ of the optimal linear upwind scheme is evaluated by
\begin{equation}
\begin{split}
&\hat{\bm{F}}_{j+1/2}^{UPS}=\frac{1}{60}\left(\bm{F}_{j-2} -8\bm{F}_{j-1}+37\bm{F}_{j}+37\bm{F}_{j+1}-8\bm{F}_{j+2}+\bm{F}_{j+3} \right)   \\
& + \sum\limits_{s=0}^{2} \bm{R}_{j+1/2}^{s} \lambda ^s \bm{L}_{j+1/2}^{s} \cdot 
\left(\bm{U}_{j-2}-5\bm{U}_{j-1}+10\bm{U}_{j}-10\bm{U}_{j+1}+5\bm{U}_{j+2}-\bm{U}_{j+3} \right). \\
\end{split}
\label{linearupwind}
\end{equation}
It is obvious that this linear scheme omits the computation of the nonlinear weights as presented in Eq.~\eqref{eq8} and ~\eqref{reconstruction},
and decreases by $5/6$ the characteristic-projection operations of Eq.~\eqref{eqchara} as noted in Ref.~\cite{hu2015efficient}.
Therefore, the hybridization will save much computational time.

The threshold is defined as
\begin{equation}
\epsilon=C\left( \frac{\Delta x}{L}\right)^{\alpha} , \\
\label{eq31}
\end{equation}
where $L$ is the characteristic length scale of the problem, $\alpha$ is a positive integer and $C$ is a positive constant.
In this paper, we choose $C=1$ and $\alpha=3$.
When $\sigma^{s}<\epsilon$, we think the flow field is smooth enough and numerical flux is evaluated by the optimal linear upwind scheme.
Otherwise, it is obtained by the regularized WENO-IS scheme.

\section{Numerical cases }
\label{sec1}
In order to assess the performance of the proposed scheme, various test cases are considered, 
including shock tube problems, shock/entropy wave interaction problems and a problem involving very strong discontinuities.  
In this section, the local Lax-Friendrichs flux is adopted for flux splitting in one dimensional cases, 
while Lax-Friendrichs flux is used for two dimensional cases if not mentioned otherwise.
The third order TVD Runge-Kutta scheme is utilized for time integration with a CFL number of 0.5.
In the following, "WENO-JS" denotes the classical 5th-order WENO scheme~\cite{jiang1996efficient}, 
"WENO-Z" represents the scheme devised in Ref.~\cite{borges2008improved}, 
"WENO-IS" denotes the original scheme in Ref.~\cite{wang2018incremental}, 
"WENO-HY" is the hybrid method devised by Hu et al.~\cite{hu2015efficient}, 
"Present" is the proposed scheme of this paper, and "Exact" denotes the theoretical or convergent solution.

\subsection{Propagation of broadband sound waves }
\label{subsec1}

This case is taken form Ref.~\cite{sun2011class}.
 A propagation of sound wave packet which contains various length scales of acoustic turbulent structures is computed.
The initial condition is given as
 \begin{equation}
 \begin{split}
 &p\left( x,0\right)=p_0\left(1+ \varepsilon  \sum\limits_{k=0}^{N/2}  
 \left( E_p\left(k \right) \right)^{1/2} sin\left( 2\pi k \left(x+ \phi_k \right) \right)    \right),     \\
 &\rho\left( x,0\right)=\rho_0\left(p\left(x,0 \right) /p_0 \right)^{1/{\gamma}}    ,\\
 &u\left( x,0\right)=u_0+\frac{2}{\gamma-1}\left(c\left( x,0\right)/c_0  \right)    ,\\
 \end{split}
 \label{wave-propagation}
 \end{equation} 
where
 \begin{equation}
E_p\left( k\right)=\left(\frac{k}{k_0} \right)^4 exp^{-2\left( k/k_0\right)^2 }       \\
\label{wave-propagation1}
\end{equation} 
is the energy spectrum which reaches its maximum at $k=k_0$. Here, $\phi_k$ is a random number 
between $0$ and $1$, $\varepsilon=0.001$, $\gamma=1.4$ and the sound speed $c=\sqrt{\gamma p/\rho}$.
The computational domain is $x\in\left[ 0,1\right] $ and period boundary conditions are imposed at the boundaries. 
The computing is executed on a 128-point grid for one period of time with CFL number of $0.2$. In this case, we choose $k_0=12$,
 which means most part of the energy is concentrated on high wavenumbers~\cite{sun2011class}. 
 The numerical result of pressure distribution is shown in Fig.~\ref{wave-propagation}. 
  We can find that the original WENO-IS scheme produces more excessive numerical dissipation for the high wavenumber waves 
  although it has better numerical stability property as stated in Ref.~\cite{wang2018incremental}. 
However, with the modification of this paper, the present method achieves better resolution properties
 than WENO-Z and WENO-JS schemes, especially in the region near critical points.
\begin{figure}
	\centering
	{\includegraphics[width=0.6\textwidth]{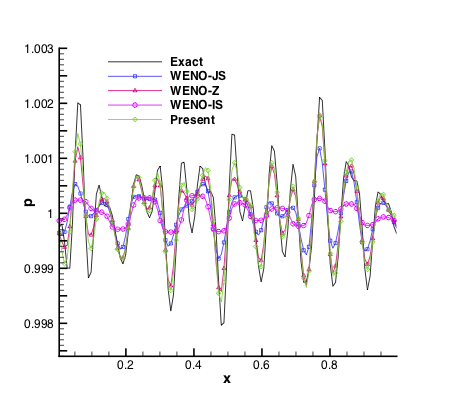}}
	\caption{ Broadband wave propagation: density profile with $k_0=12$. }
	\label{wave-propagation}
\end{figure}

\subsection{Shock-tube problems }
\label{subsec1}
In this test case, we show the proposed regularized WENO-IS scheme passing the shock tube problems, 
which include the Sod problem~\cite{sod1978survey}, the Lax problem~\cite{lax1954weak} and the 1-2-3 problem~\cite{einfeldt1991godunov}. 
These computations are executed on a 200-point grid.

The initial condition for Sod problem is
 \begin{equation}
 \left( \rho, u, p\right) =\left\lbrace  
 \begin{array}{rcl}
 \left( 1,0,1\right) &{0 \leq x<0.5}\\
 \left( 0.125,0,0.1\right) &{0.5\leq x<1}\\
 \end{array} \right.
 \label{sod}
 \end{equation}
 and the final time is $t=2$.

The initial condition for Lax problem is
\begin{equation}
\left( \rho, u, p\right) =\left\lbrace  
\begin{array}{rcl}
\left( 0.445,0.698,3.528\right) &{0 \leq x<0.5}\\
\left( 0.5,0,0.5710\right) &{0.5\leq x<1}\\
\end{array} \right.
\label{Lax}
\end{equation}
and the final time is $t=0.14$.

The initial condition for 123 problem is
\begin{equation}
\left( \rho, u, p\right) =\left\lbrace  
\begin{array}{rcl}
\left( 1,-2,0.4\right) &{0 \leq x<0.5}\\
\left( 1,2,0.4\right) &{0.5\leq x<1}\\
\end{array} \right.
\label{123}
\end{equation}
and the final time is $t=1$. 

\begin{figure}\centering
	\subfigure {\includegraphics[width=0.45\textwidth]{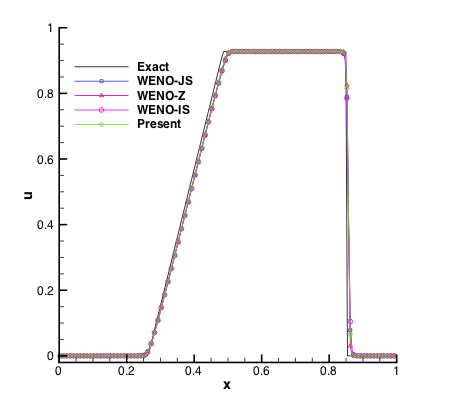}}
	\subfigure {\includegraphics[width=0.45\textwidth]{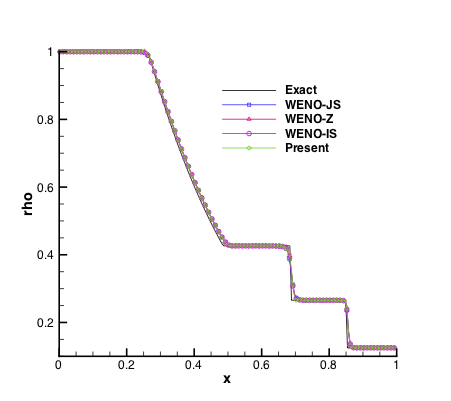}}
	\subfigure {\includegraphics[width=0.45\textwidth]{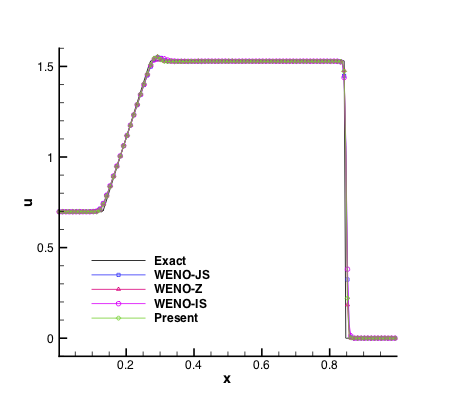}}
	\subfigure  {\includegraphics[width=0.45\textwidth]{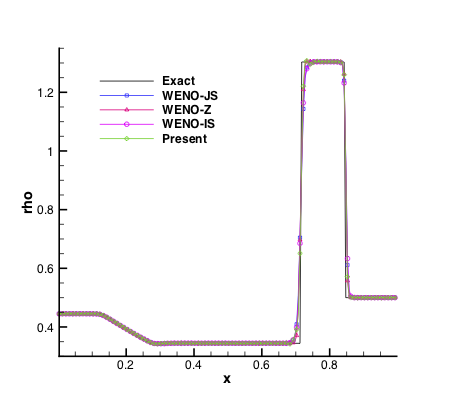}}
	\subfigure {\includegraphics[width=0.45\textwidth]{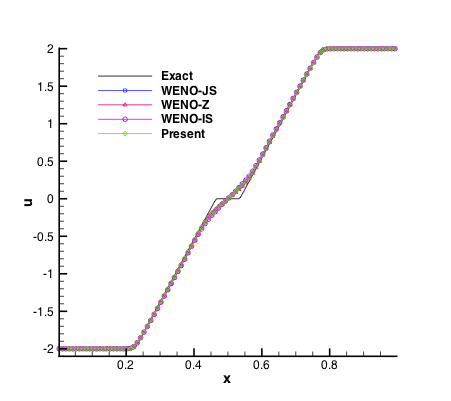}}
	\subfigure  {\includegraphics[width=0.45\textwidth]{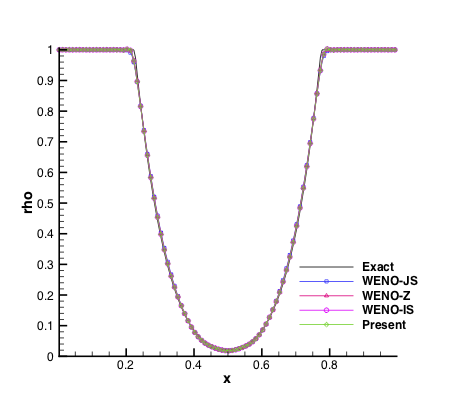}}
	\caption{  Shock tube problems: velocity and density profiles of Sod (top), Lax (middle) and 123 problems (bottom).   }
	\label{1d-shock tube}
\end{figure}

The comparison of the numerical results of density and velocity distribution is shown in Fig.~\ref{1d-shock tube}. 
The "Exact" denotes the theoretical solution. 
It is obvious that the present method performs as well as other WENO schemes.

\subsection{Interaction blast waves }
\label{subsec1}
A two-blast-wave interaction problem ~\cite{woodward1984numerical} is considered and the initial condition is
\begin{equation}
\left( \rho, u, p\right) =\left\lbrace  
\begin{array}{rcl}
\left( 1,0,1000\right) &{0 \leq x<0.1}\\
\left( 1,0,0.01\right) &{0.1\leq x<0.9}\\
\left( 1,0,100\right) &{0.9\leq x<1}.\\
\end{array} \right.
\label{two-blast}
\end{equation}
The Roe flux is adopted for flux splitting and a reflection boundary condition is applied at $x=0$ and $x=1$.
The computing is carried on up to $t=0.038$ on a 400-point grid. 
\begin{figure}\centering
	\subfigure {\includegraphics[width=0.45\textwidth]{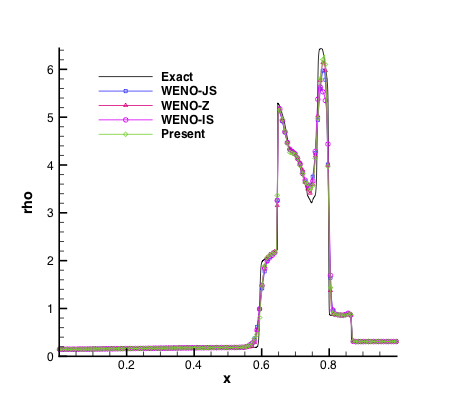}}
	\subfigure {\includegraphics[width=0.45\textwidth]{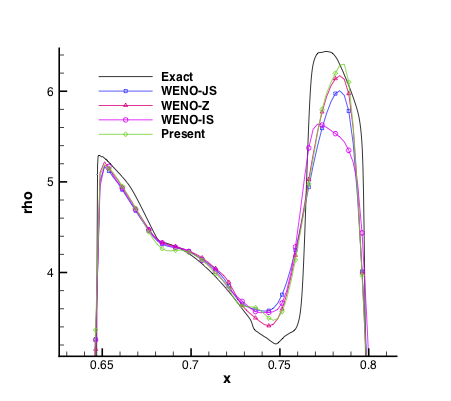}}
	\caption{  Interacting blast waves: density profile (left) and a zoom of the density profile (right).   }
	\label{blast-wave}
\end{figure}
Fig.~\ref{blast-wave} gives the comparison of density profile and a enlarged part. 
The "Exact" denotes the convergent result which is computed by WENO-JS at the resolution of $N=2500$.
It can be observed that all methods capture the strong shock waves very well.
However, the original WENO-IS scheme show considerate larger numerical dissipation and the present method exhibits comparatively less dissipation than others, 
especially near the right peak of the density profile as shown in Fig.~\ref{blast-wave}.

\subsection{The Shu-osher problem}
\label{subsec1}
This case gives a Mach 3 shock wave interaction with a sine entropy wave and the initial condition is 
\begin{equation}
\left( \rho, u, p\right) =\left\lbrace  
\begin{array}{rcl}
\left( 3.857143,2.629369,10.3333\right) &{x\leq 1}\\
\left( 1+0.2sin\left(5x \right) ,0,1\right) &{otherwise}.\\
\end{array} \right.
\label{shu-osher}
\end{equation}
The computing is carried on a 200-point grid with the domain of $x\in\left[0,10 \right] $ and the finial time is $t=1.8$.
The density profiles computed by different schemes are shown in Fig.~\ref{shu-osher}. 
The reference "Exact" denotes a high-resolution result solved by WENO-JS on a 3200-point grid.
It can be clearly noticed that all schemes can capture the shock waves well.
However, in the region downstream the shock wave, the present method shows superior resolution 
than WENO-Z, WENO-JS and WENO-IS schemes in reproducing the fine flow structures.
Also, compared with TENO5 and TENO5-opt scheme as shown in ~\cite{fu2016family}(their Fig. 15),
the density profile of the present method is more close to the convergent density profile especially in the region downstream the shock wave.
\begin{figure}\centering
	\subfigure {\includegraphics[width=0.45\textwidth]{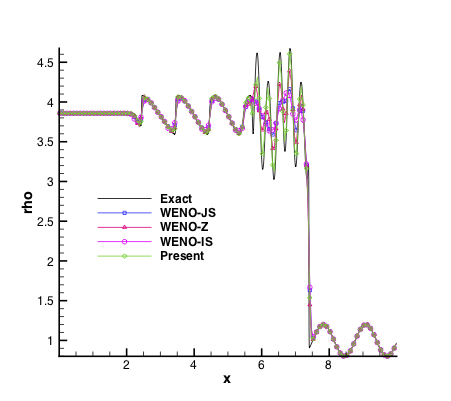}}
	\subfigure {\includegraphics[width=0.45\textwidth]{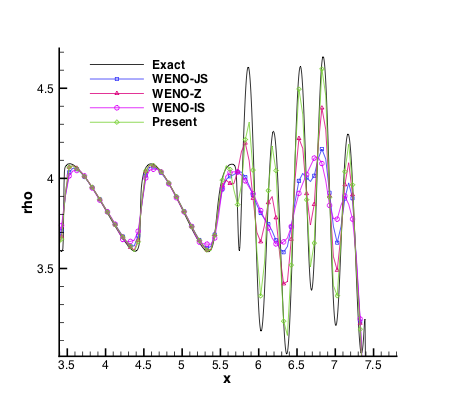}}
	\caption{  Shu-osher problem: density profile (left) and a zoom of the density profile (right).   }
	\label{shu-osher}
\end{figure}

\subsection{Double Mach reflection of a strong shock }
\label{subsec1}
This is a two-dimensional test case taken from Woodward and Colella~\cite{woodward1984numerical} 
on the double Mach reflection of a strong shock.
The initial condition is 
\begin{equation}
\left( \rho, u,v, p\right) =\left\lbrace  
\begin{array}{rcl}
\left( 1.4,0,0,1.0\right)\quad  &{y\leq 1.732\left(x-0.1667 \right) }\\
\left( 8.0,7.145,-4.125,116.8333\right) &{otherwise},\\
\end{array} \right.
\label{double-mach}
\end{equation}
and the computation is solved on the domain of $\left[ 0,4\right] \times \left[0,1 \right]  $.
 This case represents a right moving shock wave with Mach 10 initially locates at $x=0.1667$, $y=0$ 
 and makes an angle of $60^{\circ}$
with $x-$axis. For the lower boundary, the reflection wall condition is imposed for $0.1667<x<4$, 
while the post shock condition is applied for $0<x<0.1667$.
The computation is carried on two grids with solutions of $512\times 128$ and $1024\times 256$ and the final time is $t=0.2$.

\begin{figure}\centering
	\subfigure {\includegraphics[width=0.48\textwidth]{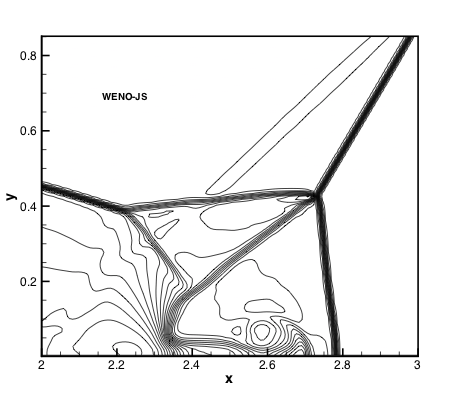}}
	\subfigure {\includegraphics[width=0.48\textwidth]{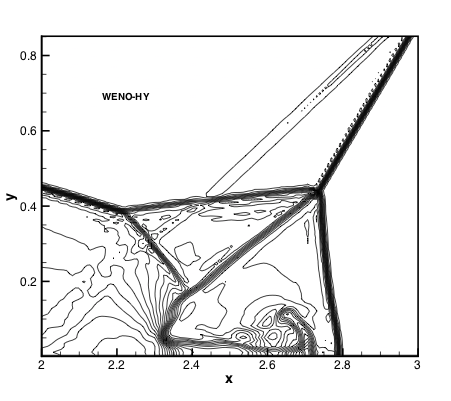}}
	\subfigure {\includegraphics[width=0.48\textwidth]{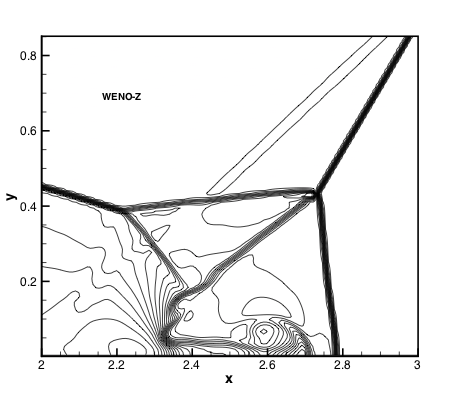}}
	\subfigure {\includegraphics[width=0.48\textwidth]{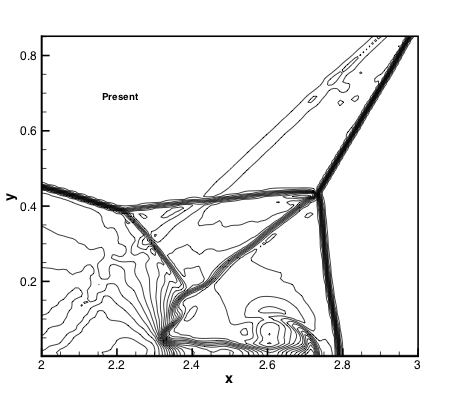}}
	\caption{  Density profiles of double Mach reflection of a strong shock on the $512\times 128$ grid. 
		43 density contours between 1.887 and 20.9 are shown in this figure.   }
	\label{double-mach512}
\end{figure}

\begin{figure}\centering
	\subfigure {\includegraphics[width=0.48\textwidth]{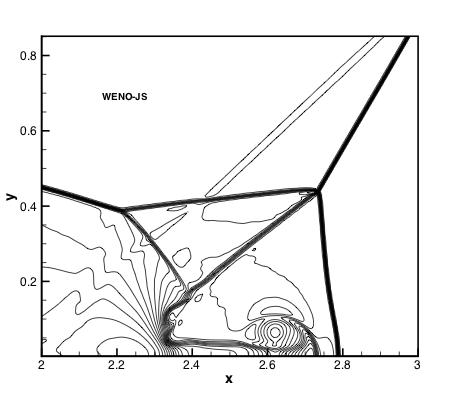}}
    \subfigure {\includegraphics[width=0.48\textwidth]{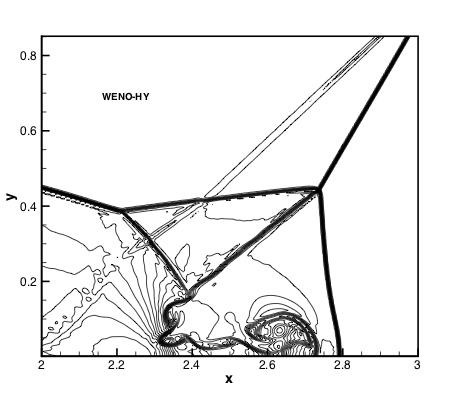}}
    \subfigure {\includegraphics[width=0.48\textwidth]{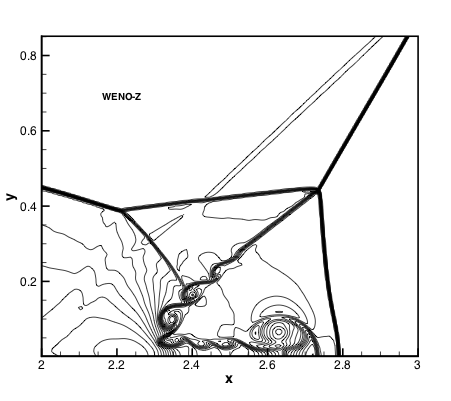}}
    \subfigure {\includegraphics[width=0.48\textwidth]{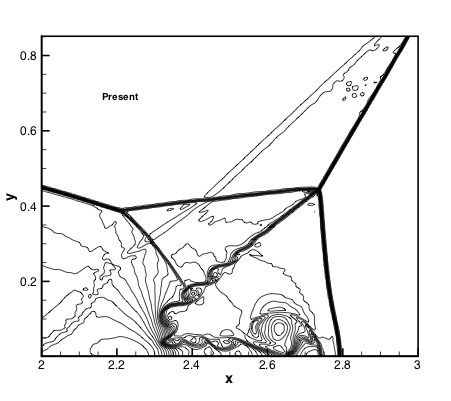}}
	\caption{  Density profiles of double Mach reflection of a strong shock on the $1024\times 256$ grid. 
		43 density contours between 1.887 and 20.9 are shown in this figure.   }
	\label{double-mach1028}
\end{figure}

The density profiles on $512\times 128$ and $1024\times 256$ grids 
are shown in Fig.~\ref{double-mach512} and Fig.~\ref{double-mach1028}, respectively.
The close-up view of the "blow up" region is presented to compare the fine flow structure resolving abilities of different schemes.
It can be noticed that all methods can capture the main flow features, such as the Mach stem and the near wall jet. 
However, the fine structures around the slip line and the near wall jet are considerably different in their solutions. 
Among these schemes, the present method resolves the finest flow structures and strongest near wall jet.
Although WENO-Z scheme resolves nearly the same amount of roll-up structures with the present method,
it is inferior in terms of the direction and the shape of the near wall jet compared with 
the results with very high resolutions, e.g. Fig.8 in Ref.~\cite{han2011wavelet}, 
Fig.2.1 in Ref.~\cite{shi2003resolution} and Fig.20 in Ref.~\cite{gerolymos2009very}.
Table~\ref{double-mach-time} gives the CPU time for the entire computing, which counts from the initialization to output, 
of double Mach reflection problem  with different methods on the $512\times 128$ grid.
 We can find that the present method is computational efficient
 and only costs less than three fourth of WENO-Z.
 
\begin{table}
	\scriptsize
	\centering
	\caption{Computational time for double Mach reflection of a strong shock on the $512\times 128$ grid.}
	\begin{tabularx}{13.5cm}{@{\extracolsep{\fill}}llllr}
		\hline
		\quad & WENO-JS &  WENO-HY &  WENO-Z&  Present\\
		\hline
		CPU time (s) & $1721$	& $1146$& $1986$& $1446$ \\
		\hline
	\end{tabularx}
	\label{double-mach-time}
\end{table}

\subsection{Two-dimensional viscous shock-tube problem }
\label{subsec1}
This viscous shock tube problem is taken from Ref.~\cite{daru2000evaluation} and 
has been investigated in Refs.~\cite{sjogreen2003grid,zhou2018grid}.
In this problem, the shock wave reflected from the end wall interacts with the boundary layer 
on the side wall induced by the incident shock. 
These interactions result in a complex system of vortices, shock wave bifurcation and other various flow structures. 
The flow filed containing various fine structures makes it a perfect case for testing high-order schemes.
The initial condition is
\begin{equation}
\left( \rho, u,v, p\right) =\left\lbrace  
\begin{array}{rcl}
\left( 120,0,0,120/\gamma    \right)  &{0\leq x<1/2 }\\
\left( 1.2,0,0,1.2/\gamma    \right) &{1/2\leq x<1},\\
\end{array} \right.
\label{viscous shock-tube}
\end{equation}
where $\gamma =1.4$ with ideal gas used. The non-slip adiabatic condition is applied to all boundary conditions of the tube.
Since the configuration is symmetric about the line $y=0.5$, only half of the domain is computed, 
which is $\left[0,1 \right] \times \left[0,0.5 \right]  $.
Here, we set the Prandtl number $Pr=0.73$ and the final time is $1.0$.
The viscosity is assumed to be constant and the Reynolds number of 200 and 1000 are considered. 

\begin{figure}\centering
	\subfigure {\includegraphics[width=0.48\textwidth]{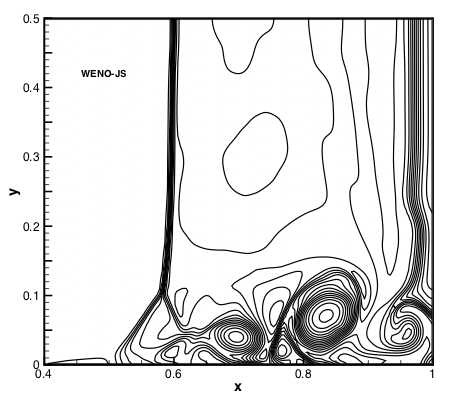}}
	\subfigure {\includegraphics[width=0.48\textwidth]{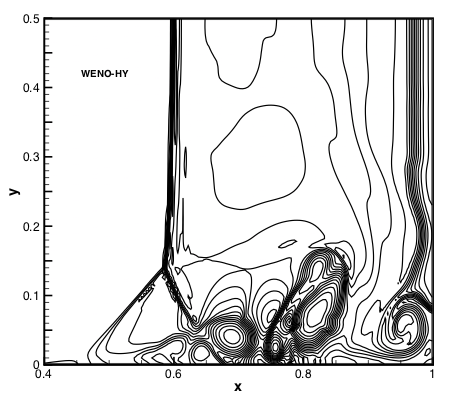}}
	\subfigure {\includegraphics[width=0.48\textwidth]{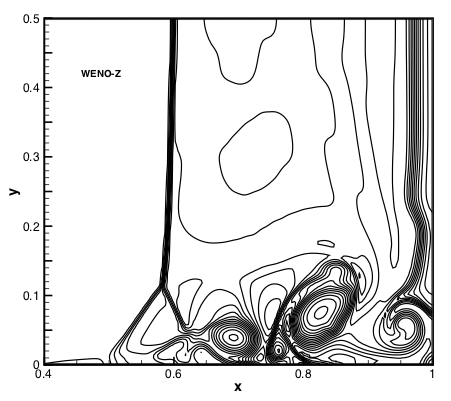}}
	\subfigure {\includegraphics[width=0.48\textwidth]{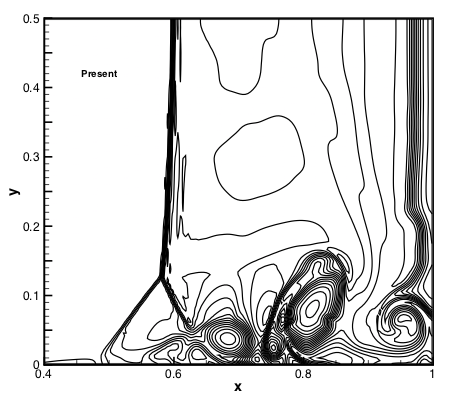}}
	\caption{ Viscous shock tube problem: 20 density contours between $15$ and $125$. This result is solved on the $250\times125$ grid.  }
	\label{viscous-shock-250}
\end{figure}

The density profiles of the problem with the Reynolds number of 200 are shown in Fig.~\ref{viscous-shock-250}.
The grid converged density profiles in Sj{\"o}green and Yee~\cite{sjogreen2003grid}(their Fig.2) and 
Zhou et al.~\cite{zhou2018grid}(their Fig.3) obtained with much higher grid resolutions
are taken as the reference for comparison.
Considering the position of the triple point, the shape and the height of the primary vortex and the orientation of the long axis of the primary vortex, 
it is obvious that the present method gives considerable better result than other schemes. 
The comparison of the entire computational time of this case by various methods is shown in Table ~\ref{viscous-tube-250}. 
We can see that the present method is more efficient than WENO-JS and WENO-Z scheme and only costs nearly three fourth of WENO-Z in this case. 

\begin{figure}\centering
	\subfigure {\includegraphics[width=0.48\textwidth]{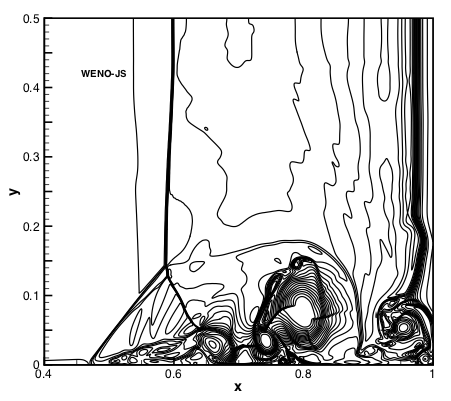}}
	\subfigure {\includegraphics[width=0.48\textwidth]{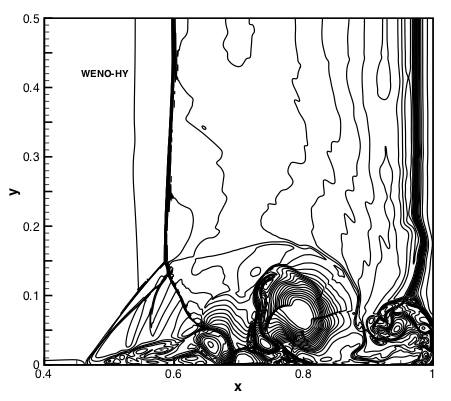}}
	\subfigure {\includegraphics[width=0.48\textwidth]{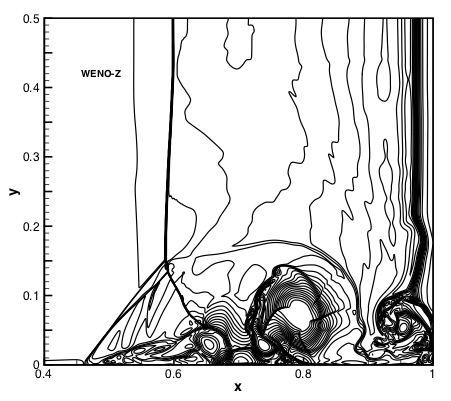}}
	\subfigure {\includegraphics[width=0.48\textwidth]{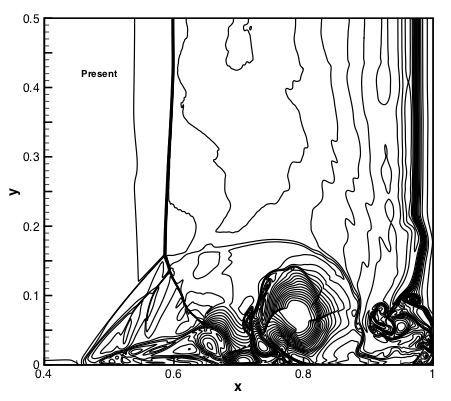}}
	\caption{ Viscous shock tube problem: 20 density contours between $20$ and $115$. This result is solved on the $1028\times640$ grid.  }
	\label{viscous-shock-1028}
\end{figure}

Fig.~\ref{viscous-shock-1028} gives the density profile of the problem with the Reynolds number of 1000.
The grid convergent result can be found in Ref.~\cite{zhou2018grid}(their Fig.6) and Ref.~\cite{fu2016family}(their Fig.21).
With the Reynolds number increased to 1000, more fine flow structures appear in the flow field.
As shown in Fig.~\ref{viscous-shock-1028}, the present method performs well in terms of shock capturing and fine-structure resolution. 
We can see that the lambda-shape shocks captured by the present method and the WENO-Z scheme are more sharply than that of WENO-JS and WENO-HY.
The big rotating structures at the lower right corner obtained by WENO-JS, WENO-HY, WENO-Z and TENO5 
(see Fig.22 in Ref.~\cite{fu2016family}) are obviously different from the reference solution.
However, the one produced by the present method fits the reference well.
Furthermore, the primary vortex, the adjacent small vortices as well as the jet beneath the oblique shock are predicted quite well with the present method.

\begin{table}
	\scriptsize
	\centering
	\caption{Computational time for viscous shock tube problem on the $250\times 125$ grid.}
	\begin{tabularx}{13.5cm}{@{\extracolsep{\fill}}llllr}
		\hline
		\quad & WENO-JS &  WENO-HY &  WENO-Z&  Present\\
		\hline
		CPU time (s) & $1690$	& $1176$& $1740$& $1343$ \\
		\hline
	\end{tabularx}
	\label{viscous-tube-250}
\end{table}

\subsection{Mach 2000 jet flows }
\label{subsec1}
\begin{figure}\centering
	\subfigure {\includegraphics[width=0.85\textwidth]{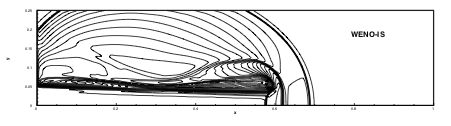}}
	\subfigure {\includegraphics[width=0.85\textwidth]{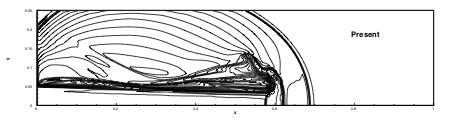}}
	\subfigure {\includegraphics[width=0.85\textwidth]{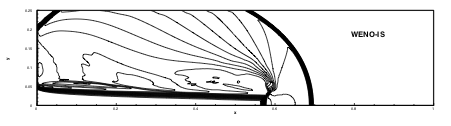}}
	\subfigure {\includegraphics[width=0.85\textwidth]{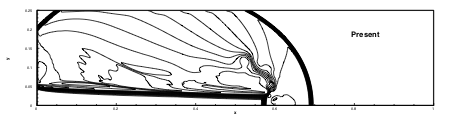}}
	\caption{  Ma 2000  jet problem: 30 density contours of logarithmic scale between $-4$ and $4$(upper); 
		30 pressure contours of logarithmic scale between $-1$ and $13$(lower). }
	\label{ma2000-iso-present}
\end{figure}

 In order to verify the outstanding numerical stability of the present method, 
 we consider the Mach 2000 jet problem, which has been computed in Zhang and Shu~\cite{zhang2010positivity} with positive preserving limiters.
The computing is carried out on the domain $\left[0,1 \right]\times \left[ 0,0.25\right]  $ with the resolution of $640\times 160$. 
The initial condition is $\left(\rho,u,v,p \right)=\left(0.5,0,0,0.4127 \right)  $ for the entire domain.
A reflective condition is applied at the bottom boundary and an outflow condition is used at the right and top boundaries.
For the left boundary, an inflow condition is adopted with states $\left(\rho,u,v,p \right)=\left(5,800,0,0.4127 \right)  $ 
when $y<0.05$ and $\left(\rho,u,v,p \right)=\left(0.5,0,0,0.4127 \right)  $ otherwise. 
We set $\gamma=5/3$ and the speed of the jet is 800, which gives about Mach 2100 with respect to the sound speed in the jet gas.
In the computing, the CFL number is set to $0.25$ and the final time is $0.001$ as in the reference~\cite{hu2013positivity}.
Note that, without additional positive preserving method applied, the WENO-JS, WENO-HY and WENO-Z schemes 
all blow out during the computing.
However, WENO-IS and the present method perform well. 
The computed density and pressure profiles of WENO-IS and the present method are shown in Fig.~\ref{ma2000-iso-present}.
We can observe that these results are in good agreement with those in Zhang and Shu~\cite{zhang2010positivity} (their Fig.4.6) 
and Hu et al.~\cite{hu2013positivity}(their Fig.4).
Furthermore, the present method shows better resolution than WENO-IS scheme and 
the classical WENO scheme combined with positive preserving flux limiters~\cite{zhang2010positivity,hu2013positivity} 
with more fine flow structures.

\section{Conclusions}
\label{sec1}
In this paper, a L2-norm regularized incremental-stencil WENO scheme is proposed for compressible flows. 
Following the stencil selection of WENO-IS scheme, we make the following modifications: 
(a) a L2-norm error term is introduced in the weighting strategy of WENO-IS scheme
to cope with the discrepancy of the smoothness indicators between 2- and 3-point stencils;
(b) a suitable non-dimensional regularization parameter is chosen to adaptively control the contribution of the error and regularized terms; 
(c) a hybridization with the optimal linear scheme is introduced to improve the performance 
of the scheme at smooth region and further improve computational efficiency.
These modifications make the method can predict more fine flow structures
by reducing numerical dissipation,
and the remaining of 2-point stencils helps to compute the flow field with strong discontinuities.
A set of test cases reveal that 
the proposed method has good fine-structure resolving capability and
keeps the superior robustness.



\section*{Acknowledgements}
\addcontentsline{toc}{section}{Acknowledgement}
The first author is partially supported by Xidian University (China) and 
the project of Natural Science Foundation of Shaanxi Province (Grant No:2019JM-186).


\bibliographystyle{elsarticle-num}

\bibliography{weno-is}

\begin{thebibliography}{10}
\expandafter\ifx\csname url\endcsname\relax
  \def\url#1{\texttt{#1}}\fi
\expandafter\ifx\csname urlprefix\endcsname\relax\def\urlprefix{URL }\fi
\expandafter\ifx\csname href\endcsname\relax
  \def\href#1#2{#2} \def\path#1{#1}\fi

\bibitem{harten1983high}
A.~Harten, High resolution schemes for hyperbolic conservation laws, Journal of
  computational physics 49~(3) (1983) 357--393.

\bibitem{harten1987uniformly}
A.~Harten, B.~Engquist, S.~Osher, S.~R. Chakravarthy, Uniformly high order
  accurate essentially non-oscillatory schemes, iii, in: Upwind and
  high-resolution schemes, Springer, 1987, pp. 218--290.

\bibitem{liu1994weighted}
X.-D. Liu, S.~Osher, T.~Chan, Weighted essentially non-oscillatory schemes,
  Journal of computational physics 115~(1) (1994) 200--212.

\bibitem{jiang1996efficient}
G.-S. Jiang, C.-W. Shu, Efficient implementation of weighted eno schemes,
  Journal of computational physics 126~(1) (1996) 202--228.

\bibitem{henrick2005mapped}
A.~K. Henrick, T.~D. Aslam, J.~M. Powers, Mapped weighted essentially
  non-oscillatory schemes: achieving optimal order near critical points,
  Journal of Computational Physics 207~(2) (2005) 542--567.

\bibitem{martin2006bandwidth}
M.~P. Mart{\'\i}n, E.~M. Taylor, M.~Wu, V.~G. Weirs, A bandwidth-optimized weno
  scheme for the effective direct numerical simulation of compressible
  turbulence, Journal of Computational Physics 220~(1) (2006) 270--289.

\bibitem{gerolymos2009very}
G.~Gerolymos, D.~S{\'e}n{\'e}chal, I.~Vallet, Very-high-order weno schemes,
  Journal of Computational Physics 228~(23) (2009) 8481--8524.

\bibitem{borges2008improved}
R.~Borges, M.~Carmona, B.~Costa, W.~S. Don, An improved weighted essentially
  non-oscillatory scheme for hyperbolic conservation laws, Journal of
  Computational Physics 227~(6) (2008) 3191--3211.

\bibitem{castro2011high}
M.~Castro, B.~Costa, W.~S. Don, High order weighted essentially non-oscillatory
  weno-z schemes for hyperbolic conservation laws, Journal of Computational
  Physics 230~(5) (2011) 1766--1792.

\bibitem{taylor2007optimization}
E.~M. Taylor, M.~Wu, M.~P. Mart{\'\i}n, Optimization of nonlinear error for
  weighted essentially non-oscillatory methods in direct numerical simulations
  of compressible turbulence, Journal of Computational Physics 223~(1) (2007)
  384--397.

\bibitem{sun2011class}
Z.-S. Sun, Y.-X. Ren, C.~Larricq, S.-y. Zhang, Y.-c. Yang, A class of finite
  difference schemes with low dispersion and controllable dissipation for dns
  of compressible turbulence, Journal of computational physics 230~(12) (2011)
  4616--4635.

\bibitem{hu2010adaptive}
X.~Hu, Q.~Wang, N.~A. Adams, An adaptive central-upwind weighted essentially
  non-oscillatory scheme, Journal of Computational Physics 229~(23) (2010)
  8952--8965.

\bibitem{hu2011scale}
X.~Hu, N.~A. Adams, Scale separation for implicit large eddy simulation,
  Journal of Computational Physics 230~(19) (2011) 7240--7249.

\bibitem{fu2016family}
L.~Fu, X.~Y. Hu, N.~A. Adams, A family of high-order targeted eno schemes for
  compressible-fluid simulations, Journal of Computational Physics 305 (2016)
  333--359.

\bibitem{hu2013positivity}
X.~Y. Hu, N.~A. Adams, C.-W. Shu, Positivity-preserving method for high-order
  conservative schemes solving compressible euler equations, Journal of
  Computational Physics 242 (2013) 169--180.

\bibitem{zhang2010positivity}
X.~Zhang, C.-W. Shu, On positivity-preserving high order discontinuous galerkin
  schemes for compressible euler equations on rectangular meshes, Journal of
  Computational Physics 229~(23) (2010) 8918--8934.

\bibitem{wang2018incremental}
B.~Wang, G.~Xiang, X.~Y. Hu, An incremental-stencil weno reconstruction for
  simulation of compressible two-phase flows, International Journal of
  Multiphase Flow 104 (2018) 20--31.

\bibitem{nielsen2015neural}
M.~A. Nielsen, Neural networks and deep learning, Vol.~25, Determination press
  USA, 2015.

\bibitem{hu2015efficient}
X.~Hu, B.~Wang, N.~A. Adams, An efficient low-dissipation hybrid weighted
  essentially non-oscillatory scheme, Journal of Computational Physics 301
  (2015) 415--424.

\bibitem{sod1978survey}
G.~A. Sod, A survey of several finite difference methods for systems of
  nonlinear hyperbolic conservation laws, Journal of computational physics
  27~(1) (1978) 1--31.

\bibitem{lax1954weak}
P.~D. Lax, Weak solutions of nonlinear hyperbolic equations and their numerical
  computation, Communications on pure and applied mathematics 7~(1) (1954)
  159--193.

\bibitem{einfeldt1991godunov}
B.~Einfeldt, C.-D. Munz, P.~L. Roe, B.~Sj{\"o}green, On godunov-type methods
  near low densities, Journal of computational physics 92~(2) (1991) 273--295.

\bibitem{woodward1984numerical}
P.~Woodward, P.~Colella, The numerical simulation of two-dimensional fluid flow
  with strong shocks, Journal of computational physics 54~(1) (1984) 115--173.

\bibitem{han2011wavelet}
L.~Han, T.~Indinger, X.~Hu, N.~A. Adams, Wavelet-based adaptive
  multi-resolution solver on heterogeneous parallel architecture for
  computational fluid dynamics, Computer Science-Research and Development
  26~(3-4) (2011) 197.

\bibitem{shi2003resolution}
J.~Shi, Y.-T. Zhang, C.-W. Shu, Resolution of high order weno schemes for
  complicated flow structures, Journal of Computational Physics 186~(2) (2003)
  690--696.

\bibitem{daru2000evaluation}
V.~Daru, C.~Tenaud, Evaluation of tvd high resolution schemes for unsteady
  viscous shocked flows, Computers \& fluids 30~(1) (2000) 89--113.

\bibitem{sjogreen2003grid}
B.~Sj{\"o}green, H.~C. Yee, Grid convergence of high order methods for
  multiscale complex unsteady viscous compressible flows, Journal of
  computational physics 185~(1) (2003) 1--26.

\bibitem{zhou2018grid}
G.~Zhou, K.~Xu, F.~Liu, Grid-converged solution and analysis of the unsteady
  viscous flow in a two-dimensional shock tube, Physics of Fluids 30~(1) (2018)
  016102.

\end{thebibliography}

\end{document}